\documentclass[aps,prl,twocolumn,superscriptaddress,showpacs,preprintnumbers,amsmath,amssymb]{revtex4}
\usepackage{graphicx} 
\usepackage{dcolumn}  
\usepackage{color}    
\usepackage{bm}       
\usepackage{ulem}     

\graphicspath{{ps}}

\def\BBbar{B\bar{B}}
\def\qqbar{q\bar{q}}

\def\Bz{B^{0}}

\def\pip{\pi^+}
\def\pim{\pi^-}
\def\piz{\pi^0}
\def\kp{K^{+}}
\def\pipi{\pip \pim}
\def\Kpi{K^{+}\pim}
\def\epem{e^+ e^-}
\def\Kstz{K^{*0}}

\def\rzKstz{\rho^0 \Kstz}
\def\fzKstz{f_0(980) \Kstz}
\def\rzKpi{\rho^0 \Kpi}
\def\fzKpi{f_0(980) \Kpi}
\def\pipiKstz{\pipi\Kstz}
\def\pipiKpi{\pipi\Kpi}
\def\ftKstz{f_2(1270) \Kstz}

\def\aonek{a^{-}_{1}(1260)\kp}

\def\YrzKstz{77.6^{+28.6}_{-27.9}}
\def\YfzKstz{51.2^{+20.4}_{-19.3}}
\def\YrzKpi{207.8^{+39.8}_{-39.2}}
\def\YfzKpi{106.9^{+31.6}_{-29.9}}
\def\YpipiKstz{200.7^{+46.7}_{-44.9}}
\def\YpipiKpi{-5.4^{+54.9}_{-44.9}}

\def\SrzKstz{2.7}
\def\SfzKstz{2.5}
\def\SrzKpi{5.0}
\def\SfzKpi{3.5}
\def\SpipiKstz{4.5}
\def\SpipiKpi{0.0}

\def\ULrzKstz{< 3.4}
\def\ULfzKstz{< 2.2}
\def\ULfzKpi{< 2.1}
\def\ULpipiKpi{< 2.1}

\def\BFrzKstz{2.1^{+0.8}_{-0.7}{^{+0.9}_{-0.5}}}
\def\BFfzKstz{1.4^{+0.6}_{-0.5}{^{+0.6}_{-0.4}}}
\def\BFrzKpi{2.8\pm0.5\pm0.5}
\def\BFfzKpi{1.4\pm0.4{^{+0.3}_{-0.4}}}
\def\BFpipiKstz{4.5^{+1.1}_{-1.0}{^{+0.9}_{-1.6}}}
\def\BFpipiKpi{-0.1^{+1.2}_{-1.1}{^{+1.4}_{-0.8}}}
\def\BFunit{\times10^{-6}}

\def\absBFrzKpi{2.8\pm0.5(\rm stat)\pm0.5(\rm syst)}

\def\Y4S{\Upsilon(4S)}
\def\BF{{\cal B}}
\def\Lz{{\cal L}_{0}}
\def\Lmax{{\cal L}_{\rm max}}

\def\NBBUsed{657 \times 10^6}

\def\GeV{{\rm GeV}}
\def\GeVc{{\rm GeV\!/}c}
\def\GeVcc{{\rm GeV\!/}c^2}

\def\Ebeam{E_{\rm beam}}
\def\DE{\Delta E}
\def\Mbc{M_{\rm bc}}
\def\Mpipi{M_{\pi\pi}}
\def\MKpi{M_{K\pi}}
\def\CosB{\cos \theta_B}
\def\Cos2B{\cos^2 \theta_B}
\def\Dz{\Delta z}
\def\Ls{{\cal L}_s}
\def\Lqq{{\cal L}_{\qqbar}}
\def\Rqq{{\cal R}_{\qqbar}}

\def\btos{b\to s}
\def\btoc{b\to c}
\def\btou{b \to s,u,d}

\renewcommand{\arraystretch}{1.1}

\begin{document}

\title{\quad\\[0.5cm] 
Measurements of Charmless Hadronic \bm{$\btos$}
  Penguin Decays in the \bm{$\pipiKpi$} Final State and First Observation of
  \bm{$B^0 \to \rho^{0}K^{+}\pi^{-}$}}

\affiliation{Budker Institute of Nuclear Physics, Novosibirsk}
\affiliation{University of Cincinnati, Cincinnati, Ohio 45221}
\affiliation{T. Ko\'{s}ciuszko Cracow University of Technology, Krakow}
\affiliation{Department of Physics, Fu Jen Catholic University, Taipei}
\affiliation{The Graduate University for Advanced Studies, Hayama}
\affiliation{Hanyang University, Seoul}
\affiliation{University of Hawaii, Honolulu, Hawaii 96822}
\affiliation{High Energy Accelerator Research Organization (KEK), Tsukuba}
\affiliation{Hiroshima Institute of Technology, Hiroshima}
\affiliation{Institute of High Energy Physics, Chinese Academy of Sciences, Beijing}
\affiliation{Institute of High Energy Physics, Vienna}
\affiliation{Institute of High Energy Physics, Protvino}
\affiliation{Institute of Mathematical Sciences, Chennai}
\affiliation{Institute for Theoretical and Experimental Physics, Moscow}
\affiliation{J. Stefan Institute, Ljubljana}
\affiliation{Kanagawa University, Yokohama}
\affiliation{Korea University, Seoul}
\affiliation{Kyungpook National University, Taegu}
\affiliation{\'Ecole Polytechnique F\'ed\'erale de Lausanne (EPFL), Lausanne}
\affiliation{Faculty of Mathematics and Physics, University of Ljubljana, Ljubljana}
\affiliation{University of Maribor, Maribor}
\affiliation{University of Melbourne, School of Physics, Victoria 3010}
\affiliation{Nagoya University, Nagoya}
\affiliation{Nara Women's University, Nara}
\affiliation{National Central University, Chung-li}
\affiliation{National United University, Miao Li}
\affiliation{Department of Physics, National Taiwan University, Taipei}
\affiliation{H. Niewodniczanski Institute of Nuclear Physics, Krakow}
\affiliation{Nippon Dental University, Niigata}
\affiliation{Niigata University, Niigata}
\affiliation{University of Nova Gorica, Nova Gorica}
\affiliation{Novosibirsk State University, Novosibirsk}
\affiliation{Osaka City University, Osaka}
\affiliation{Panjab University, Chandigarh}
\affiliation{Saga University, Saga}
\affiliation{University of Science and Technology of China, Hefei}
\affiliation{Seoul National University, Seoul}
\affiliation{Shinshu University, Nagano}
\affiliation{Sungkyunkwan University, Suwon}
\affiliation{University of Sydney, Sydney, New South Wales}
\affiliation{Tata Institute of Fundamental Research, Mumbai}
\affiliation{Toho University, Funabashi}
\affiliation{Tohoku Gakuin University, Tagajo}
\affiliation{Department of Physics, University of Tokyo, Tokyo}
\affiliation{Tokyo Metropolitan University, Tokyo}
\affiliation{Tokyo University of Agriculture and Technology, Tokyo}
\affiliation{IPNAS, Virginia Polytechnic Institute and State University, Blacksburg, Virginia 24061}
\affiliation{Yonsei University, Seoul}
  \author{S.-H.~Kyeong}\affiliation{Yonsei University, Seoul} 
  \author{Y.-J.~Kwon}\affiliation{Yonsei University, Seoul} 
  \author{I.~Adachi}\affiliation{High Energy Accelerator Research Organization (KEK), Tsukuba} 
  \author{H.~Aihara}\affiliation{Department of Physics, University of Tokyo, Tokyo} 
  \author{A.~M.~Bakich}\affiliation{University of Sydney, Sydney, New South Wales} 
  \author{V.~Balagura}\affiliation{Institute for Theoretical and Experimental Physics, Moscow} 
  \author{E.~Barberio}\affiliation{University of Melbourne, School of Physics, Victoria 3010} 
  \author{A.~Bay}\affiliation{\'Ecole Polytechnique F\'ed\'erale de Lausanne (EPFL), Lausanne} 
  \author{K.~Belous}\affiliation{Institute of High Energy Physics, Protvino} 
  \author{M.~Bischofberger}\affiliation{Nara Women's University, Nara} 
% ...
\collaboration{The Belle Collaboration}

\begin{abstract}

  We report measurements of charmless hadronic $B^{0}$ decays into the
  $\pipiKpi$ final state. The analysis uses a sample of $\NBBUsed$
  $\BBbar$ pairs collected with the Belle detector at the KEKB
  asymmetric-energy $\epem$ collider at the $\Y4S$ resonance.  The
  decay $B^0\to\rzKpi$ is observed for the first time; the
  significance is $5.0\sigma$ and the corresponding partial branching
  fraction for $\MKpi \in (0.75,1.20)~\GeVcc$ is
  $[\absBFrzKpi]\BFunit$.  We also obtain the first evidence for $B^0
  \to \fzKpi$ with $\SfzKpi\sigma $ significance and for $B^0
  \to\pipiKstz$ with $\SpipiKstz\sigma$ significance.  For the
  two-body decays $B^0\to\rzKstz$ and $B^0\to\fzKstz$, the
  significances are $2.7\sigma$ and $2.5\sigma$, respectively, and the
  upper limits on the branching fractions are $3.4\BFunit$ and
  $2.2\BFunit$ at 90\% confidence level.

\end{abstract}

\pacs{11.30.Er, 13.25.Hw, 14.40.Nd}


{\maketitle


{\renewcommand{\thefootnote}{\fnsymbol{footnote}}}
\setcounter{footnote}{0}  

In the Standard Model (SM), charmless hadronic $B$
meson decays occur mainly via two processes: (i) $b \to s q \bar{q}$
transitions mediated by penguin diagrams, and (ii) $b \to u W^*$
transitions mediated by tree diagrams.  These diagrams as they
pertain to $B^0 \to \rho^0 K^{*0}$~\cite{conjugate}, for example,
are shown in Fig.~\ref{Feyn}.  Both of these processes are
suppressed relative to the more common $b \to c W^*$ decays due to
either (i) the one-loop structure or (ii) the small ratio of CKM
matrix~\cite{CKM} elements $|V_{ub}/V_{cb}|$, respectively.  Because
of this suppression, these decays are especially sensitive to non-SM
contributions~\cite{nonSM}.

\begin{figure}[b]
  \begin{tabular}{ccc}
    \includegraphics[width=0.22\textwidth]{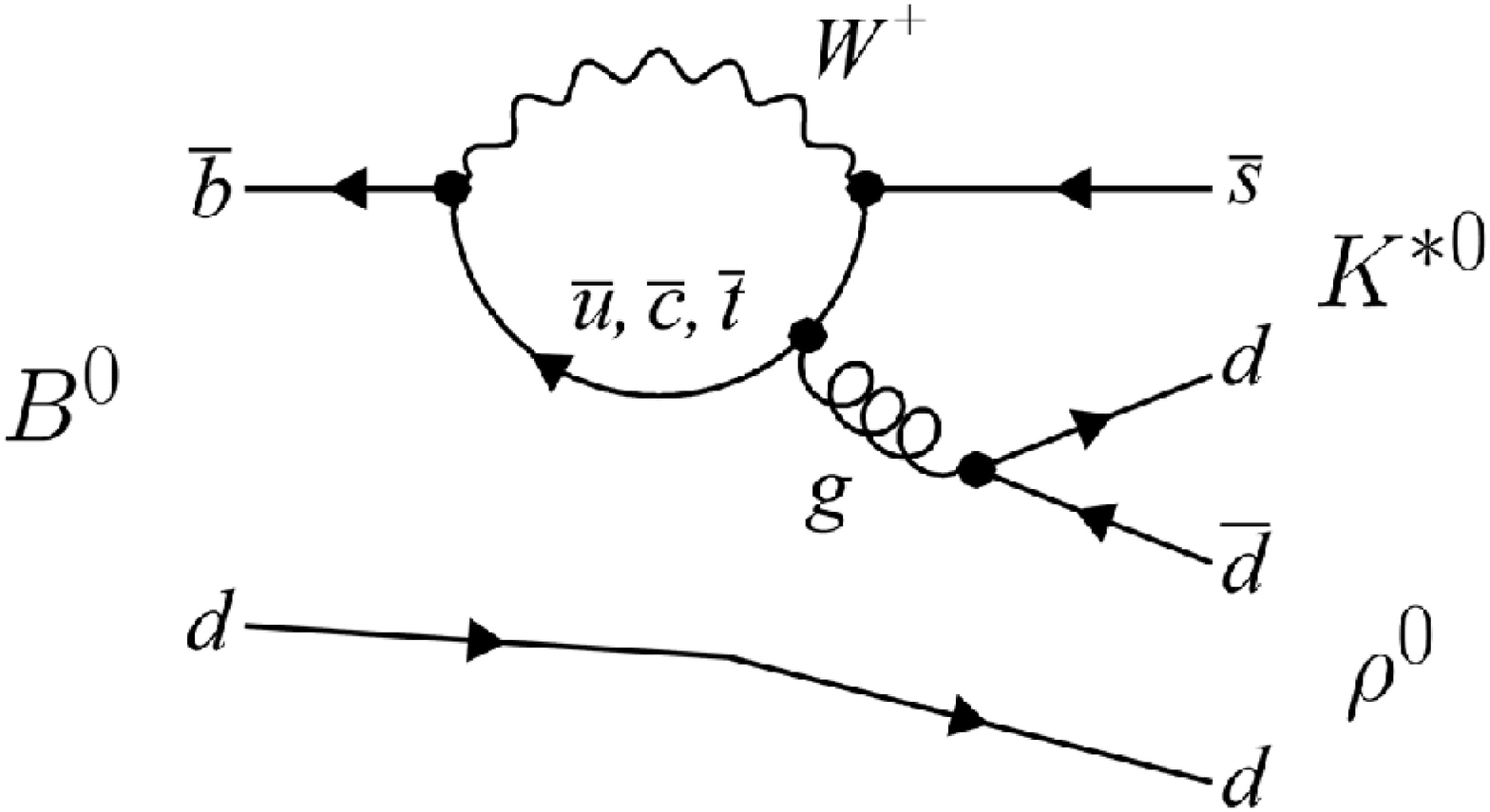} &&
    \includegraphics[width=0.22\textwidth]{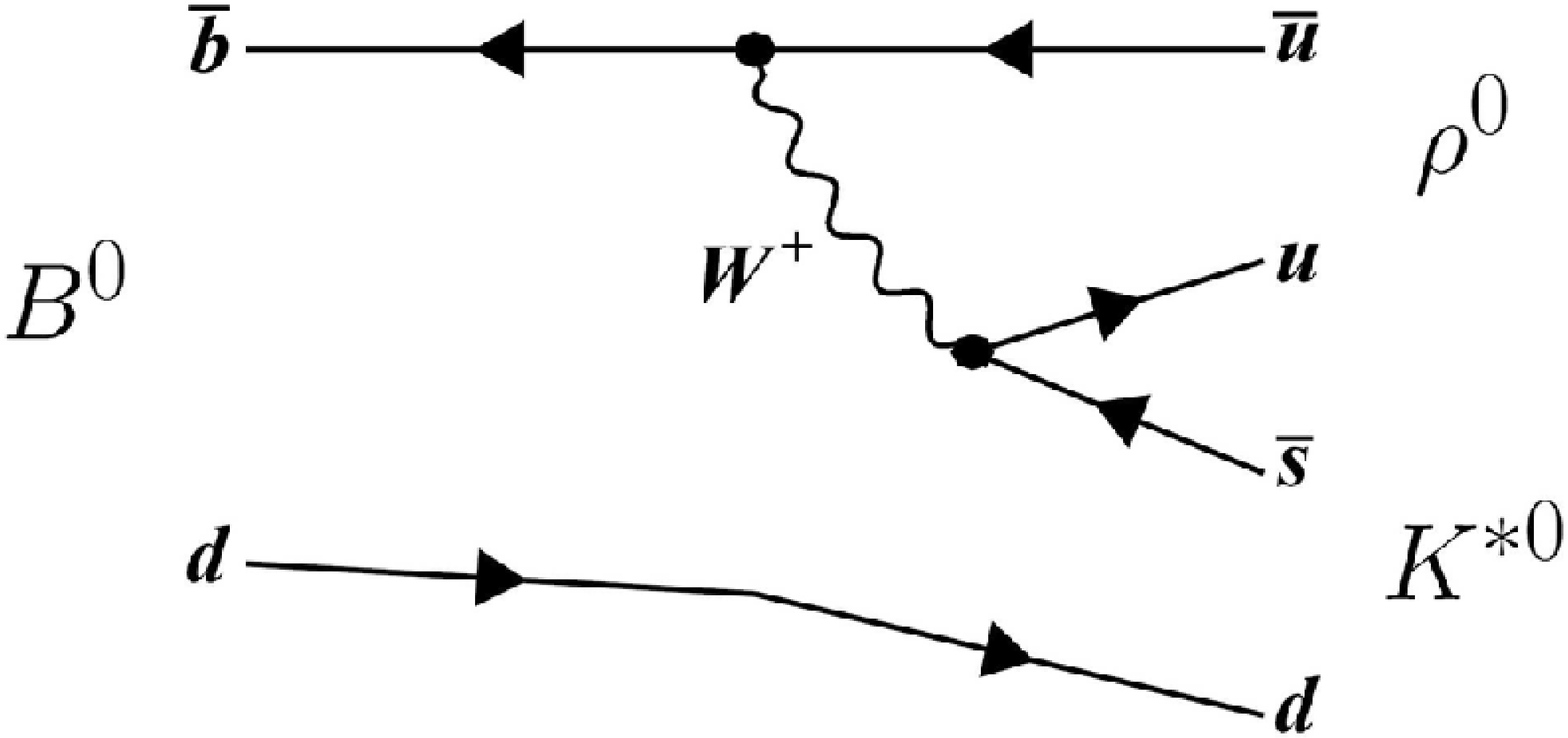}\\
    (a) &\hspace{0.17cm} & (b)
  \end{tabular}
  \caption{Feynman diagrams for charmless hadronic $B$ decays
    pertaining to $\Bz\to\rzKstz$: (a) $\btos$ penguin diagram,
    (b) $b\to u$ tree diagram. }
  \label{Feyn}
\end{figure}

There have been several puzzling results from
measurements of charmless hadronic $B$ decays.  For example, $B$
decays to $K^+ \pim$ and $K^+ \piz$ show different patterns of direct
$CP$ violation~\cite{Nature}, which are inconsistent with na\"{\i}ve SM
expectations.  It has been suggested~\cite{shed} that vector-vector
($VV$) final states with the same quark combinations, e.g. $B \to \rho
K^*$  may give insights to this puzzle, as any
difference between $K\pi$ and their $VV$ counterparts will be mainly
hadronic.  In addition, charmless $B$ decays to $VV$ final states show
intriguing results in the final-state polarizations.
The decays $B \to \phi K^*$ and $B \to \rho K^*$,
both occurring mostly via the $b\to s$ penguin process, are found to have
large transverse polarizations~\cite{expB2phiKst, rzKstzBaBar,
  rpKstzBelle}, in contrast to the expectation from factorization.  On
the other hand, $B^{+(0)} \to \rho^+ \rho^{0(-)}$,
which is mostly a $b\to u W^*$ tree-diagram process,
is almost fully polarized
longitudinally~\cite{rhorho}.  There have been
theoretical~\cite{shed,rzKst:theory} studies of these modes, in part
focusing on the final-state polarizations within and beyond the SM.

One difficulty in measuring charmless $B \to VV$
decays, however, is that non-resonant decays to the same final
state can be a significant background~\cite{nonResVV}.  Such
non-resonant decays may have different decay properties, e.g.,
different polarization of the vector mesons.  While
there are several experimental studies of $B^0$ decays to the
$\pipiKpi$ final
state~\cite{rzKstzBaBar, exp_pipiKpi}, there is no
experimental information on the non-resonant components of these
final states.

In this paper, we analyze charmless hadronic decays of $B^0$ to the
$\pipiKpi$ final state. We search for two-body
final states such as $B^0 \to \rzKstz$ and $\fzKstz$, and also for
three-body states $\rzKpi $, $\fzKpi$~\cite{fz_BF},
and $\pipiKstz$, where the $\pipi$  or $\Kpi$
pairs are non-resonant.
A comprehensive understanding of these decays with
a clear distinction between
non-resonant and two-body resonant decays
would advance our understanding of strong and weak interaction
dynamics.

We use a data sample containing $\NBBUsed$ $\BBbar$ pair events
collected with the Belle detector~\cite{Belle} at the KEKB~\cite{KEKB}
asymmetric-energy $\epem$ collider ($3.5$ on $8~\GeV$), operating at
the $\Upsilon(4S)$ resonance. 
To reconstruct $\Bz$ $\to$ $\pipiKpi$ decays including the
intermediate states $\rho^0$ $\to$ $\pi^+\pi^-$, $f_0(980)$ $\to$
$\pi^+\pi^-$, and $\Kstz\to\Kpi$, we select four charged tracks of
which two are positively charged and two are negatively charged. Each
track is required to originate within 5.0~cm of the interaction
point (IP) along the beam direction, and within 0.2~cm of the IP in
the transverse plane~\cite{rpKstzBelle}.  We also require that the 
transverse momentum of each track be larger than 
$0.1~\GeVc$~\cite{rpKstzBelle}.  Tracks identified as electrons
are rejected.  We identify charged kaons and pions by
combining particle identification (PID) information obtained from the
central drift chamber, the time-of-flight system, 
and the aerogel Cherenkov
counters~\cite{PID}.

Signal candidates are selected for further analysis based on four
kinematic variables: the $\pi^+\pi^-$ and $K^+\pi^-$ invariant masses
($\Mpipi$ and $\MKpi$), the energy difference $\DE \equiv E_B -
\Ebeam$, and the beam-energy-constrained mass $\Mbc \equiv
\sqrt{\Ebeam^2 - p_B^2}$, where $\Ebeam$ is the beam energy and $E_B$
and $p_B$ are the energy and momentum, respectively, of the candidate
$B$-meson.  These variables are all evaluated in the $\Y4S$
center-of-mass (CM) frame. We retain events satisfying $|\DE| <
0.1~\GeV$, $5.24~\GeVcc < \Mbc < 5.29~\GeVcc$, $0.55~\GeVcc < \Mpipi <
1.20~\GeVcc$ and $0.75~\GeVcc < \MKpi < 1.20~\GeVcc$.  To optimize the
background suppression criteria, tighter ``signal regions'' are
defined for $\Mbc$ and $\DE$: $5.27~\GeVcc < \Mbc < 5.29~\GeVcc$ and
$|\DE|<0.045~\GeV$.  The fraction of events having multiple candidates
is approximately 20\%. For multiple-candidate events, we select the
candidate decay having the smallest $\chi^2$ from the $B$ vertex fit.
Given a set of four particles, $\pipiKpi$, two combinations of ($\pi^+
\pi^-$) and ($K^+ \pi^-$) may lie inside the selected mass ranges.  In
this case, which occurs in less than 1\% of signal decays, we pair the
higher-momentum $\pi^-$ with the $\pi^+$.

The dominant source of background is from continuum $\epem \to \qqbar$
events ($q = u,\ d,\ s,\ c$). These events are distinguished from the
signal by their event shape.  Since $B$ mesons are spinless and
produced nearly at rest in the CM frame, their daughter particles are
distributed almost isotropically.  On the other hand, continuum events
usually produce two back-to-back jets in the CM frame.  We use Monte
Carlo (MC) simulated~\cite{evtgen} signal events and sideband data
($5.20~\GeVcc < \Mbc < 5.26~\GeVcc $) for optimizing the
continuum suppression requirements. First we form a Fisher discriminant
${\cal F}$ based on a set of modified Fox-Wolfram
moments~\cite{KSFW}. These moments are uncorrelated with the four
kinematical variables mentioned above.  Two more variables are used
for continuum suppression: $\CosB$, the cosine of the polar angle of
the $B$ flight direction in the CM frame; and $\Dz$, the displacement
along the beam direction between the vertex of the signal $B$ and that
of the other $B$ in the event.  Likelihood functions for signal
($\Ls$) and continuum background ($\Lqq$) are formed from products of
the probability density functions (PDFs) for ${\cal F}$, $\CosB$, and
$\Dz$.  These are combined into a likelihood ratio $\Rqq = \Ls / (\Ls
+ \Lqq)$.  To obtain improved continuum suppression, we optimize the
requirement on $\Rqq$ as a function of flavor-tagging information from
the accompanying $B$ meson.  The Belle flavor-tagging
algorithm~\cite{TaggingNIM} yields the $b$-flavor variable $q$ (=$\pm
1$), and the quality variable $r$.  The latter ranges from zero for no
flavor discrimination to one for unambiguous flavor assignment.  
We optimize the $\Rqq$ requirement independently in six bins of $qr$. For
$B^0 \to \rzKstz$, for example, the optimized $\Rqq$ requirements
remove 99\% of the $\qqbar$ background while retaining 42\% of the
signal.

$B$ decays to a charm meson ($D^0$ or $D^{(*)+}$) and multiple pions
constitute a significant background that exhibits peaking behavior in
$\Mbc$ and $\DE$ similar to that of the signal.  To eliminate this
background, we veto candidates that have a $K\pi\pi$, $K\pi$, or
$\pi\pi$ invariant mass consistent with a $D^{(*)+} \to K^- \pip\pip$,
$D^0 \to K^- \pip$, and $D^0 \to \pim\pip$ decay, respectively.

The signal yields are obtained from a four-dimensional extended
unbinned maximum-likelihood~\cite{LYONS} fit (4D fit) to $\Mbc$, $\DE$, $\Mpipi$
and $\MKpi$. The likelihood function is
\begin{eqnarray}
\label{eqLikeFtn}
{\cal L} & \equiv &
\frac{\exp(-\sum Y_j)}{N!} \prod_{i=1}^{N} \sum_{j} Y_j {\cal P}_{~j}^{i}\,,
\end{eqnarray}
where $Y_j$ is the yield of the $j$-th component, ${\cal P}_{~j}^i$ is
the PDF value for the $j$-th component of the $i$-th event, and $i$
runs over all events in the fit region ($N$).  We include 13
components in Eq.~(\ref{eqLikeFtn}): $B^0$ decays to $\rzKstz$,
$\fzKstz$, and $f_2(1270)K^{*0}$; the non-resonant components
$\rzKpi$, $\fzKpi$, $\pipi \Kstz$, and $\pipiKpi$; the feed-down
components $\aonek$, $K^{+}_{1}(1270) \pi^-$, and $K^{+}_{1}(1400)
\pi^-$; and background components from $\qqbar$ continuum ($\qqbar$),
charmed $B$-decays ($b\to c$), and charmless $B$-decays ($\btou$).

The PDFs for the signal are separated into two categories: correctly
reconstructed events and self-cross-feed (SCF) events.  The SCF events
include at least one track that is taken from the accompanying $B$
meson decay.  For correctly reconstructed events, a sum of two
Gaussians with a common mean is used for the $\Mbc$ and $\DE$
shapes. The $\Mpipi$ and $\MKpi$ distributions are modeled by
relativistic Breit-Wigner functions. The $\rho^{0}$, $f_2(1270)$ and
$K^{*0}$ resonance parameters are fixed to their PDG
values~\cite{PDG}. Parameters of the $f_0(980)$ resonance shape are
fixed to the results of Ref.~\cite{MORI}; these values have higher
precision than the corresponding PDG values.  PDFs for the SCF
components are modeled using Kernel Estimation~\cite{Kernel} of SCF MC
distributions.  For the $\Mpipi$ and $\MKpi$ PDFs of non-resonant
components, a threshold function and/or Chebyshev polynomials are
used.  The $\Mbc$ and $\DE$ shapes for the signal PDFs are calibrated
using a large $B^0 \to D^-\pi^{+}$, $D^-\to K^+ \pim\pim$ control
sample, to take into account small differences observed between
MC-simulated events and data.

The PDF shapes of the $\qqbar$ background are modeled with an
ARGUS~\cite{argus_ftn} function for $\Mbc$, linear functions for
$\DE$, and combinatorial shapes for $\Mpipi$ and $\MKpi$.  For $b\to
c$ background, the PDFs are obtained separately for correctly
reconstructed $K^{*0}$ and for random $K\pi$ combinations.  The
fraction of each component is fixed from the MC simulation.  The PDF
shapes for $\btou$ background are modeled with non-parametric PDFs
using Kernel Estimation~\cite{Kernel}.

\begin{figure}[t]
  \includegraphics[width=0.48\textwidth]{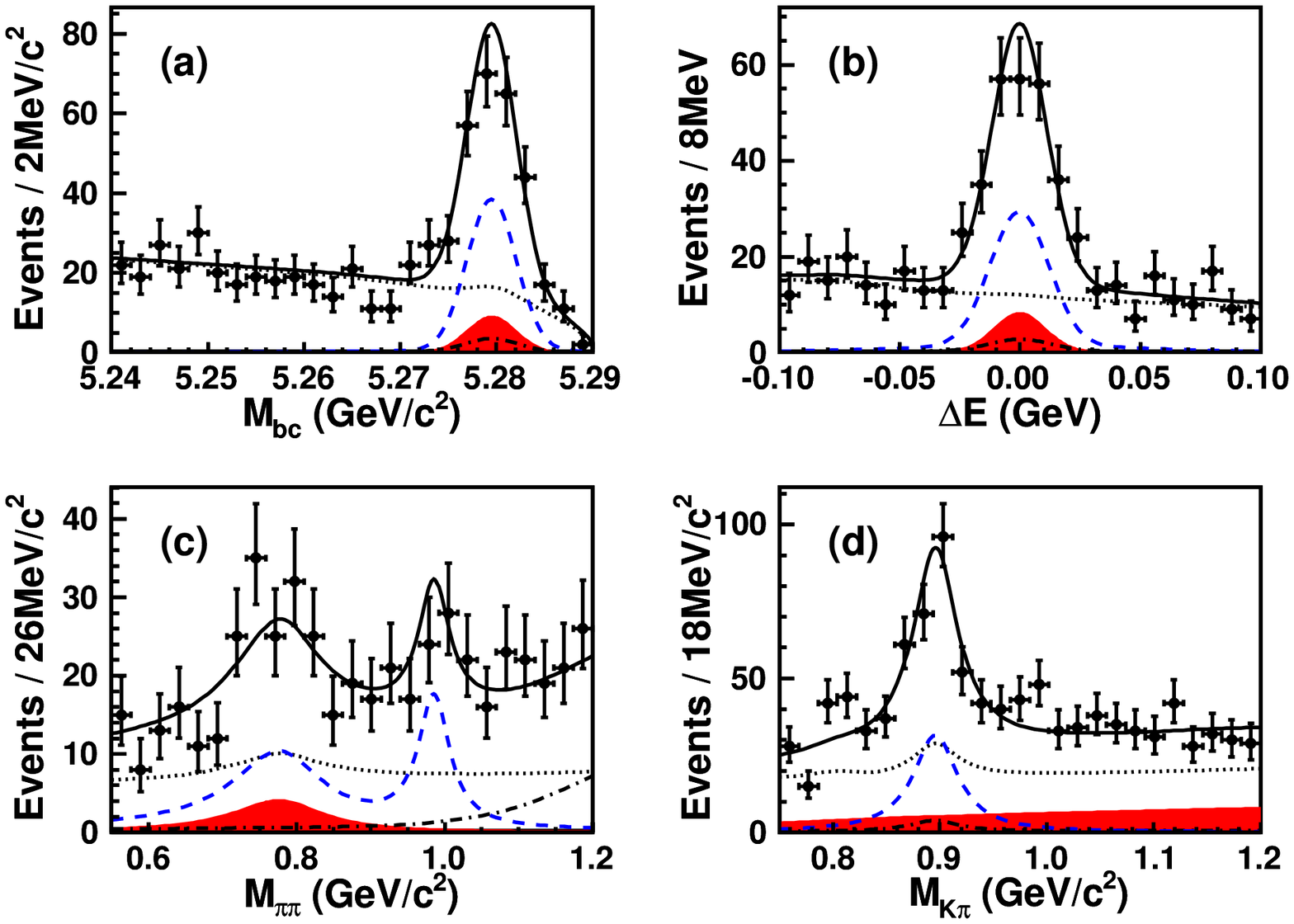}
  \caption{(color online) Projection of the 4D fit results onto (a) $\Mbc$, (b)
    $\DE$, (c) $\Mpipi$, (d) $\MKpi$,
    with the other variables required to satisfy (except for
    the variable plotted) 
    $\Mbc\in(5.27,5.29)$ $\GeVcc$, $\DE\in(-0.045,0.045)$ $\GeV$,
    $\Mpipi\in(0.62,1.04)$ $\GeVcc$ and $\MKpi\in(0.84,0.94)$ $\GeVcc$.
    The curves are for the $\rzKpi$ (solid-shaded),
    the sum of $\rzKstz$ and $\fzKstz$ (dashed), 
    $\ftKstz$ and the sum of feed-down modes (dot-dashed), 
    the sum of the backgrounds (dotted), and the total (solid). 
  }
  \label{figResults}
\end{figure}

The following parameters are floated in the 4D fit: the yields of the
signal modes (given in Table~\ref{tableResults}) and background yields
of $\btoc$ and $\qqbar$; the parameters of the $\qqbar$ PDF describing
the $\Mbc$, $\DE$ and combinatorial shapes of $\Mpipi$ and
$\MKpi$. The branching fractions of the feed-down components are fixed
to the results of Ref.~\cite{feeddown_BABAR}. The yield of
$f_2(1270)K^{*0}$ is fixed to 43.0 events as obtained from
two-dimensional $\Mbc$-$\DE$ fitting in bins of $\Mpipi$ as discussed
later.  The remaining parameters are fixed to values obtained from
 MC simulations.

The fit projections are shown in Fig.~\ref{figResults}, and 
the results are summarized in Table~\ref{tableResults}.   
There are moderate correlations between some modes, which we check
by fitting an ensemble of \textsc{Geant}-simulated
 MC samples. 
We find a negligible effect on the measured signal yields.
The branching fraction of each mode is determined by ${\cal B} = Y /
(\varepsilon_{\rm MC} \varepsilon_{\rm PID}N_{B\bar{B}})$, where $Y$
is the fitted signal yield, $\varepsilon_{\rm MC}$ is the event
selection efficiency including daughter branching fractions as
obtained from MC simulation and $\varepsilon_{\rm PID}$ is an
efficiency correction ($\varepsilon_{\rm PID}$=0.96) for PID that
accounts for small differences between MC and data.  The production
rates of $\Bz\bar{B}^0$ and $B^+B^-$ pairs are assumed to be equal.

The fit yields the first observation for $B^0\to\rzKpi$ with a
significance of $\SrzKpi\sigma$.  The significance is defined as
$\sqrt{-2\ln(\Lz /\Lmax)}$, where $\Lz$ ($\Lmax$) is the value of the
likelihood function when the yield is fixed to zero (allowed to vary).
We include systematic uncertainties by smearing the likelihood
function with a Gaussian whose width is equal to the systematic
uncertainty (discussed below).  We also find evidence
for $B^0\to\fzKpi$ with a significance of $\SfzKpi\sigma$, and
evidence for $B^0\to\pipiKstz$ with a significance of
$\SpipiKstz\sigma$.  For $B^0\to\rzKstz$ and $B^0\to\fzKstz$, we
observe excesses of events with significances of $2.7\sigma$ and
$2.5\sigma$, respectively.  For the non-resonant decay components, the
${\cal B}$ and $\varepsilon_{\rm MC}$ values correspond to the ranges
$\MKpi\in(0.75,1.20)$~$\GeVcc$, $\Mpipi\in(0.55,1.20)$~$\GeVcc$, and
assume three-body phase space distributions. For modes with less than
$3\sigma$ significance, we also list a 90\% confidence level (C.L.)
upper limit.  This limit is determined via
\begin{equation}
  \frac{\int^{{\cal B}_{\rm UL}}_{0}{\cal L(\cal B)}d{\cal B}}{\int^{\infty}_{0}{\cal
      L(\cal B)}d{\cal B}}=90\%\,.
\end{equation}

\begin{table}[t]
  \renewcommand{\arraystretch}{1.3}
  \caption{The signal yield $Y$ and its statistical uncertainty,
    corrected MC efficiency $\varepsilon$ (assuming $f_{L}=0.5$ for
    $B^0\to\rzKstz$), significance ${\cal S}$ including
    the systematic uncertainties, measured branching fraction
    $\BF$ and the upper limit (UL) at the 90\% confidence level
    $\BF_{\rm UL}$. For non-resonant decay components, $\varepsilon$,
    $\BF$  and $\BF_{\rm UL}$ are obtained for $\MKpi\in(0.75,1.20)$
    $\GeVcc$ and $\Mpipi\in(0.55,1.20)$ $\GeVcc$ assuming phase space
    distributions. For the branching fraction, the first (second)
    uncertainty is statistical (systematic).}  \label{tableResults}
  \begin{ruledtabular}
  \begin{tabular}{lccccc}
    Mode& $Y$ & $\varepsilon$ & ${\cal S}$ & $\BF$& $\BF_{\rm UL}$\\
    & (events) & (\%) & ($\sigma$) & ($10^{-6}$) & ($10^{-6}$)\\\hline
    $\rzKstz$ & $\YrzKstz$&5.73 & $\SrzKstz$& $\BFrzKstz$ & $\ULrzKstz$\\
    $\fzKstz$ & $\YfzKstz$&5.56 & $\SfzKstz$& $\BFfzKstz$ & $\ULfzKstz$\\
    $\rzKpi$  & $\YrzKpi$ &11.15 & $\SrzKpi$& $\BFrzKpi$  & -\\
    $\fzKpi$  & $\YfzKpi$ &11.43 & $\SfzKpi$& $\BFfzKpi$  & $\ULfzKpi$\\
    $\pipiKstz$&$\YpipiKstz$&6.74& $\SpipiKstz$& $\BFpipiKstz$ & -\\
    $\pipiKpi$ &$\YpipiKpi$&6.84 & $\SpipiKpi$& $\BFpipiKpi$ & $\ULpipiKpi$
  \end{tabular}
  \end{ruledtabular}
\end{table}

\begin{table}[b]
  \caption{Summary of systematic uncertainties (\%) in
      the  efficiency ($\varepsilon$) determination.
    }  \label{syst_multi}
  \begin{ruledtabular}
  \begin{tabular}{lcccccc}
    Source& $\rho K^*$ & $f K^*$ & $\rho K\pi$ & $f K\pi$& $\pi\pi K^*$ & $\pi\pi K\pi$\\\hline
    MC statistics & $\pm0.5$ & $\pm0.7$ & $\pm1.3$ & $\pm1.7$ & $\pm1.3$ & $\pm2.1$\\
    Tracking       & $\pm4.2$ & $\pm4.2$ & $\pm4.2$ & $\pm4.2$ & $\pm4.2$ & $\pm4.2$\\
    PID           & $\pm3.7$ & $\pm3.7$ & $\pm3.7$ & $\pm3.8$ & $\pm3.8$ & $\pm3.7$\\
    $\Rqq$ cut    & $\pm3.4$ & $\pm3.4$ & $\pm3.4$ & $\pm3.4$ & $\pm3.4$ & $\pm3.4$\\
    $N_{B\bar{B}}$& $\pm1.4$ & $\pm1.4$ & $\pm1.4$ & $\pm1.4$ & $\pm1.4$ & $\pm1.4$\\
    $f_{\rm L}$   & $^{+16.7}_{-18.9}$& - &-  &- & - &-\\\hline
    Sum & $^{+18.0}_{-20.1}$ & $\pm6.7$ & $\pm6.8$&
    $\pm7.0$ & $\pm6.9$ & $\pm7.0$
  \end{tabular}
  \end{ruledtabular}
\end{table}

The sources and sizes of systematic uncertainties in
the efficiency determination and the yield extraction are
summarized in Tables~\ref{syst_multi} and \ref{syst_add}, respectively.
The main sources of
efficiency uncertainties are tracking (4.2\%), PID (3.7\%--3.8\%),
MC sample statistics (0.5\%--2.1\%)  and the ${\cal R}_{\qqbar}$ 
requirement (3.4\%). 
Table~\ref{syst_multi} also includes the uncertainty from
$N_{\BBbar}$ (1.4\%). While this does not affect the
efficiency determination, it leads to a
  multiplicative uncertainty in ${\cal B}$.
An additional uncertainty in the efficiency for $B^0\to\rzKstz$
arises from the unknown fraction of longitudinal polarization ($f_{\rm L}$).
For our central value, we take $f_{\rm L}$ = 0.5 and estimate the
uncertainty by considering the two extreme cases $f_{\rm L} = 0$
and $f_{\rm L} =1$.
The systematic uncertainties in the yield extraction 
are obtained by varying all fixed parameters of the PDFs 
by $\pm 1\sigma$,  feed-down yields by $\pm3\sigma$, and
the fractions of SCF and $\btou$ backgrounds by $\pm50$\%,
respectively.
We consider the effects of higher $K^{*0}$ resonances
by including a PDF for $\Bz\to\rho^0 K^*_0(1430)^0$ 
and repeating the 4D fit with its yield floated 
by extending the fitting region in $\MKpi$ to 
$1.5~\GeVcc$; the resulting changes are included
as a systematic uncertainty.

\begin{table}[t]
  \renewcommand{\arraystretch}{1.3}
  \caption{Summary of systematic uncertainties (events)
    in the signal yield ($Y$) extraction.
    }  \label{syst_add}
  \begin{ruledtabular}
  \begin{tabular}{lcccccc}
    Source& $\rho K^*$ & $f K^*$ & $\rho K\pi$ & $f K\pi$& $\pi\pi K^*$ & $\pi\pi K\pi$\\\hline
    Fitting PDFs&$^{+4.4}_{-5.4}$&$^{+12.7}_{-11.8}$&$^{+5.8}_{-9.1}$&$^{+24.1}_{-23.6}$&$^{+18.1}_{-17.4}$&$^{+29.4}_{-27.9}$\\
    $f_{\ftKstz}$&$^{+11.0}_{-11.3}$&$^{+5.9}_{-6.4}$&$^{+0.3}_{-0.3}$&$^{+0.3}_{-0.1}$&$^{+13.9}_{-13.7}$&$^{+30.0}_{-35.4}$\\
    $f_\textrm{feed-down}$&$^{+0.6}_{-1.4}$&$^{+0.1}_{-0.1}$&$^{+4.7}_{-1.5}$&$^{+0.3}_{-0.4}$&$^{+8.7}_{-3.8}$&$^{+3.2}_{-1.9}$\\
    $f_{\btou}$   &$^{+1.9}_{-2.1}$&$^{+0.1}_{-0.0}$&$^{+7.0}_{-9.8}$&$^{+0.3}_{-0.4}$&$^{+0.0}_{-1.2}$&$^{+3.7}_{-0.8}$\\
    $f_{\rm SCF}$ &$^{+2.1}_{-2.1}$&$^{+1.2}_{-1.2}$&$^{+19.9}_{-20.6}$&$^{+7.4}_{-7.3}$&$^{+8.2}_{-8.3}$&$^{+11.8}_{-11.4}$\\
    $K^{*}_{0}(1430)^0$&$^{+29.0}_{-0.0}$&$^{+14.7}_{-0.0}$&$^{+16.9}_{-12.4}$&$^{+0.0}_{-19.3}$&$^{+0.0}_{-54.8}$&$^{+69.1}_{-0.0}$\\
    Fitting bias &$^{+2.7}_{-0.0}$&$^{+4.9}_{-0.0}$&$^{+11.2}_{-0.0}$&$^{+0.0}_{-10.2}$&$^{+0.0}_{-26.6}$&$^{+0.0}_{-29.9}$\\
    Interference &$^{+6.6}_{-5.6}$&$^{+2.3}_{-0.9}$&$^{+14.7}_{-17.3}$&$^{+4.3}_{-0.0}$&$^{+3.8}_{-3.6}$& -\\\hline
    Sum & $^{+31.5}_{-12.2}$&$^{+20.5}_{-12.3}$&$^{+34.8}_{-31.3}$&$^{+25.6}_{-32.9}$&$^{+35.5}_{-69.8}$&$^{+76.2}_{-42.6}$
  \end{tabular}
  \end{ruledtabular}
\end{table}

\begin{figure}[b]
  \centering
  \includegraphics[width=0.23\textwidth]{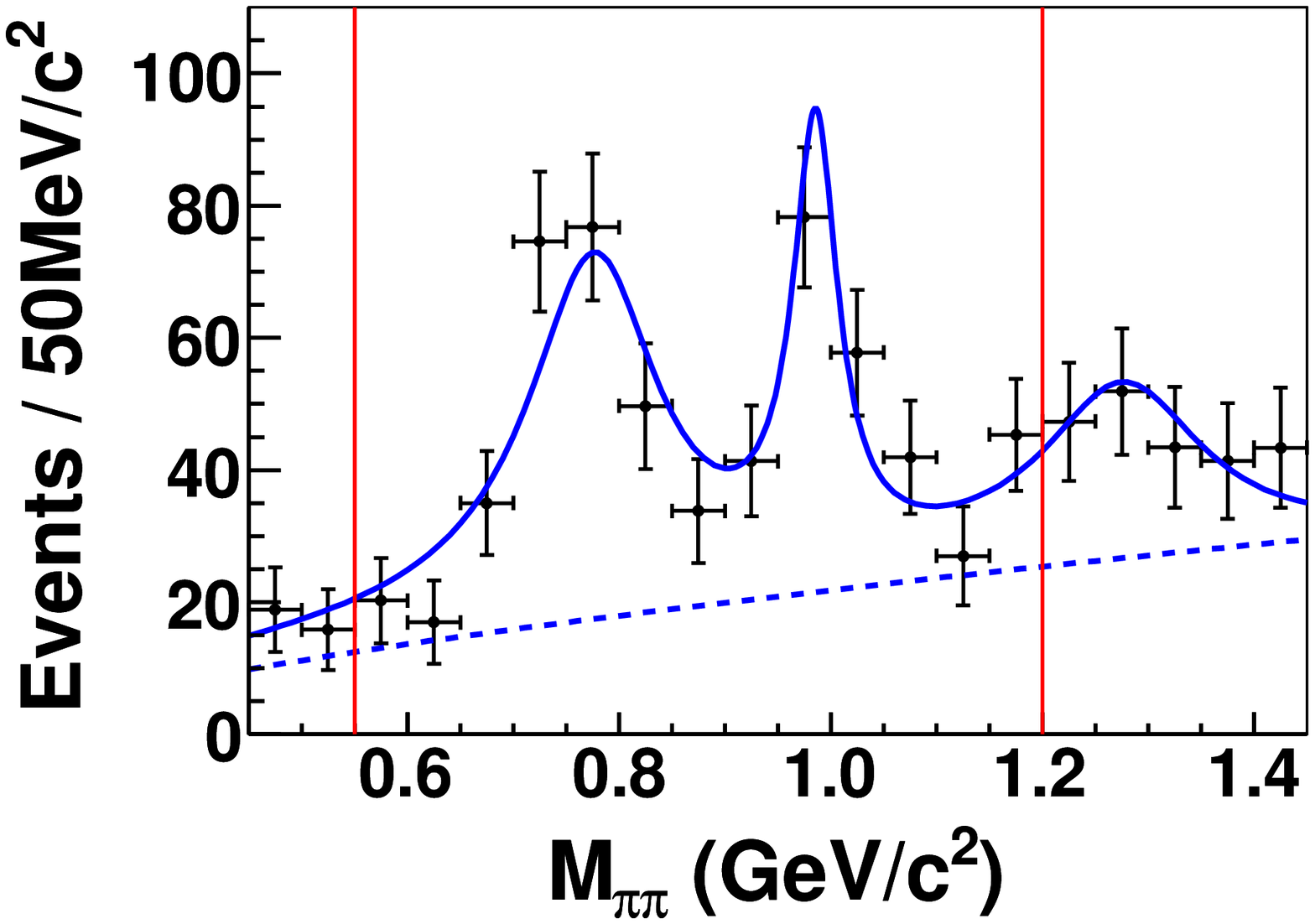}
  \includegraphics[width=0.23\textwidth]{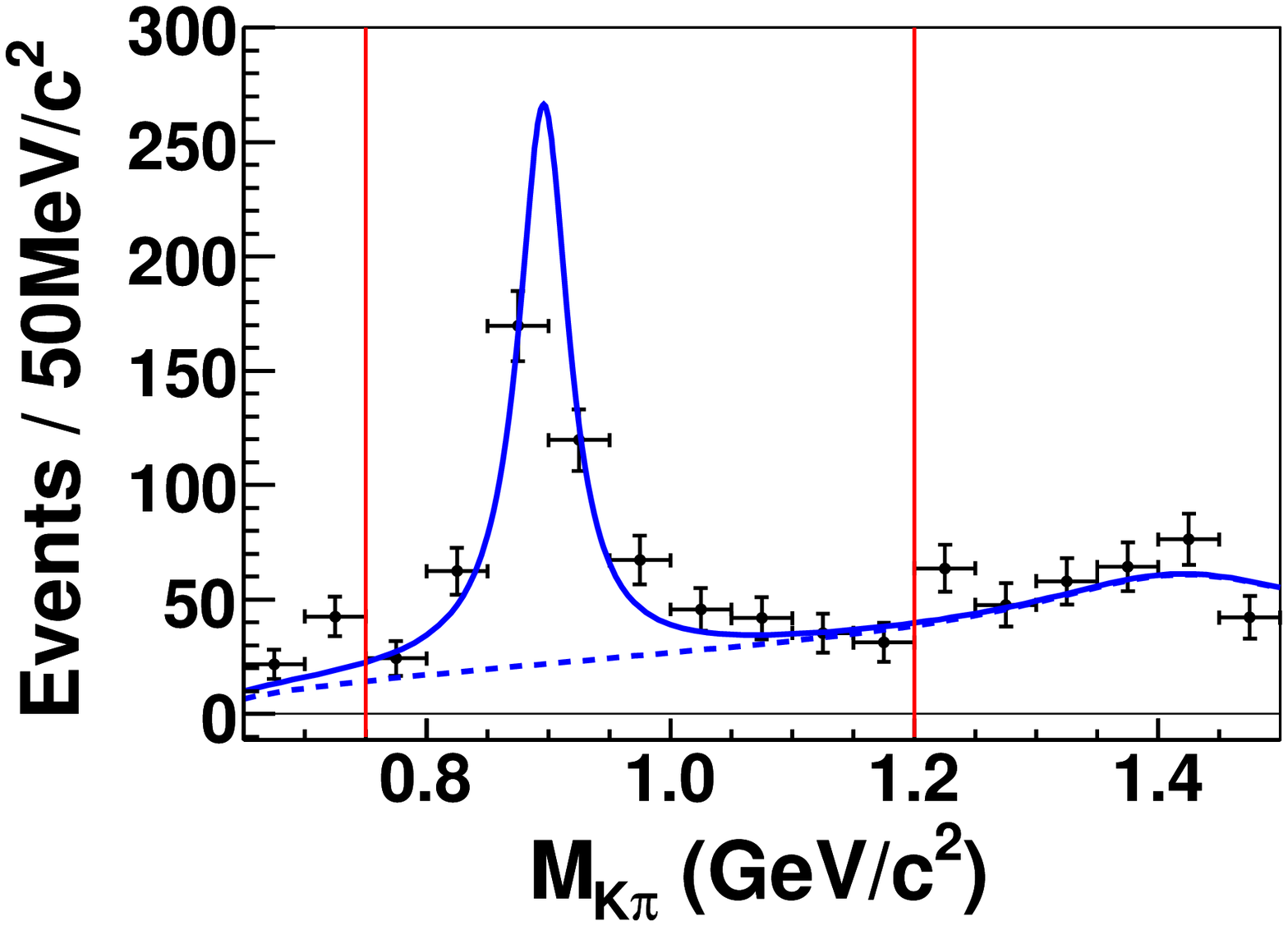}
  \centering
  \caption{Signal yields obtained from the two-dimensional fits to $\Mbc$
    and $\DE$ in bins of $\Mpipi$ (left) and $\MKpi$ (right) up to the
    higher-mass regions. Solid curves show the results of the
    two-dimensional binned fit, and dashed curves show the contributions of
    non-resonant $\pipi$ (left) and the sum of non-resonant $\Kpi$
    and $K^{*}_{0}(1430)^0$ (right). The vertical lines show the
    nominal 4D fit regions.}
  \label{bkgdsubFit}
\end{figure}

We study the effects of possible interference among $\rho^0$,
$f_0(980)$, $f_2(1270)$, and non-resonant $\pipi$ modes by including
interference terms with variable phases in the $\Mpipi$ relativistic
Breit-Wigner function.  The effect is estimated by refitting with this
modified PDF; the resulting shifts in the yields are included in the
systematic uncertainties.  We obtain the systematic
uncertainty due to possible interference between
$\Kstz$, $K^{*}_{0}(1430)^0$, and non-resonant $\Kpi$ in the $\MKpi$
mass spectrum in a similar manner.  Uncertainties due to possible
fitting bias are determined using a large sample of MC-simulated
events.  We assign the small biases found in the MC simulation as
systematic uncertainties.

To verify the large contribution from non-resonant components (see
Table~\ref{tableResults}), we study background-subtracted $\Mpipi$ and
$\MKpi$ spectra.  These spectra are obtained by binning the data in
$\Mpipi$ or $\MKpi$ and, for each bin, fitting the two-dimensional
$\Mbc$-$\DE$ distribution to determine the sum of resonant and
non-resonant yields.  Figure~\ref{bkgdsubFit} shows these yields as a
function of $\Mpipi$ and $\MKpi$.  Relativistic Breit-Wigner functions
are used as PDF's for the resonances with their parameters fixed to
their PDG values~\cite{PDG}.  The PDFs for the non-resonant
contributions are modeled by threshold functions using MC-simulated
events.

In summary, we have made the first observation of the
three-body decay $B^0\to\rzKpi$ with $\SrzKpi\sigma$ significance and
obtained the first evidence for non-resonant $B^0\to\fzKpi$
and $B^0\to\pipiKstz$ decays.
The corresponding partial branching
fractions are measured.  For the $\Bz\to\rzKstz$ and $\Bz\to\fzKstz$
modes, we find approximately $2.6\sigma$ signal
excesses and obtain the results listed in Table~\ref{tableResults}. 
Our result for these
two-body decays are $2\sigma$ and $1\sigma$ lower, respectively, than
in the previous measurement~\cite{rzKstzBaBar}.  
We have also searched for the fully non-resonant four-body decay
$\Bz\to\pipiKpi$ and calculated a 90\% C.L. upper limit on
its partial branching fraction.
Our results for the non-resonant modes are the first such results
and may help us understand the polarization puzzle in $\rho K^*$
decays. With additional $B\to VV$ data, these measurements can be
used to constrain models of new physics~\cite{beneke_2}.

We thank the KEKB group for excellent operation of the
accelerator, the KEK cryogenics group for efficient solenoid
operations, and the KEK computer group and
the NII for valuable computing and SINET3 network support.
We acknowledge support from MEXT, JSPS and Nagoya's TLPRC (Japan);
ARC and DIISR (Australia); NSFC (China);
DST (India); MEST, KISTI, and NRF (Korea); MNiSW (Poland);
MES and RFAAE (Russia); ARRS (Slovenia); SNSF (Switzerland);
NSC and MOE (Taiwan); and DOE (USA). S.-H.~K. acknowledges
support by the Seoul Science Fellowship.

\clearpage



\begin{thebibliography}{99}
\bibitem{conjugate} Throughout this paper, charge-conjugate processes
  are implied unless explicitly stated otherwise.
\bibitem{fz_BF} In this paper, we assume a branching fraction
  of 100\% for $f_0(980)$ $\to$ $\pi^+\pi^-$.
\end{thebibliography}
\end{document}